\documentclass{agujournal2019}
\usepackage{url} %this package should fix any errors with URLs in refs.
\usepackage{lineno}
\usepackage[inline]{trackchanges} %for better track changes. finalnew option will compile document with changes incorporated.
\usepackage{soul}
\usepackage{graphicx}
\usepackage{booktabs, tabularx}
\usepackage{adjustbox}
\usepackage{subcaption} 
\usepackage{multirow}
\usepackage{longtable}
\usepackage{multicol}
\usepackage{natbib}
\usepackage{apacite}
\usepackage{hyperref}
\usepackage{cleveref}
\newcommand{\rev}[1]{\textcolor{black}{\normalfont{#1}}} 
\draftfalse

\newcommand{\aap}{    {\it Astron. Astrophys. }}

\newcommand{\apjl}{   {\it Astrophys. J. Lett. }}
\newcommand{\apjs}{   {\it Astrophys. J. Suppl. S.}}

\usepackage{verbatim}

\journalname{Space Weather}

\begin{document}

%% ------------------------------------------------------------------------ %%
%  Title
%
% (A title should be specific, informative, and brief. Use
% abbreviations only if they are defined in the abstract. Titles that
% start with general keywords then specific terms are optimized in
% searches)
%
%% ------------------------------------------------------------------------ %%

% Example: \title{This is a test title}

\title{Predicting CMEs using ELEvoHI with STEREO-HI beacon data}

%% ------------------------------------------------------------------------ %%
%
%  AUTHORS AND AFFILIATIONS
%
%% ------------------------------------------------------------------------ %%

% Authors are individuals who have significantly contributed to the
% research and preparation of the article. Group authors are allowed, if
% each author in the group is separately identified in an appendix.)

% List authors by first name or initial followed by last name and
% separated by commas. Use \affil{} to number affiliations, and
% \thanks{} for author notes.
% Additional author notes should be indicated with \thanks{} (for
% example, for current addresses).

% Example: \authors{A. B. Author\affil{1}\thanks{Current address, Antartica}, B. C. Author\affil{2,3}, and D. E.
% Author\affil{3,4}\thanks{Also funded by Monsanto.}}
%TC:ignore
\authors{Maike Bauer\affil{1, 2}, Tanja Amerstorfer\affil{1}, J\"urgen Hinterreiter\affil{1, 2}, Andreas J. Weiss\affil{1, 2}, Jackie A. Davies\affil{3}, Christian M\"ostl\affil{1}, Ute V. Amerstorfer\affil{1}, Martin A.~Reiss\affil{1}, Richard A. Harrison\affil{3}}

% \affiliation{1}{First Affiliation}
% \affiliation{2}{Second Affiliation}
% \affiliation{3}{Third Affiliation}
% \affiliation{4}{Fourth Affiliation}

\affiliation{1}{Space Research Institute, Austrian Academy of Sciences, Schmiedlstraße 6, 8042 Graz, Austria}

\affiliation{2}{Institute of Physics, University of Graz, Universit\"atsplatz 5, 8010 Graz, Austria}

\affiliation{3}{RAL Space, STFC Rutherford Appleton Laboratory, Didcot, UK}
%TC:endignore
%(repeat as many times as is necessary)

%% Corresponding Author:
% Corresponding author mailing address and e-mail address:

% (include name and email addresses of the corresponding author.  More
% than one corresponding author is allowed in this LaTeX file and for
% publication; but only one corresponding author is allowed in our
% editorial system.)

% Example: \correspondingauthor{First and Last Name}{email@address.edu}

\correspondingauthor{Maike Bauer}{maike.bauer@oeaw.ac.at}

%% Keypoints, final entry on title page.

%  List up to three key points (at least one is required)
%  Key Points summarize the main points and conclusions of the article
%  Each must be 140 characters or fewer with no special characters or punctuation and must be complete sentences

% Example:
% \begin{keypoints}
% \item	List up to three key points (at least one is required)
% \item	Key Points summarize the main points and conclusions of the article
% \item	Each must be 140 characters or fewer with no special characters or punctuation and must be complete sentences
% \end{keypoints}

\begin{keypoints}
\item The viability of using the ELEvoHI model with lower-quality real-time data was studied
\item We find that using real-time data with ELEvoHI is possible, but \rev{CME} prediction benefits significantly from improved data quality
\item We find that using real-time data with ELEvoHI is possible, but coronal mass ejection prediction benefits significantly from improved data quality
\end{keypoints}

%% ------------------------------------------------------------------------ %%
%
%  ABSTRACT and PLAIN LANGUAGE SUMMARY
%
% A good Abstract will begin with a short description of the problem
% being addressed, briefly describe the new data or analyses, then
% briefly states the main conclusion(s) and how they are supported and
% uncertainties.

% The Plain Language Summary should be written for a broad audience,
% including journalists and the science-interested public, that will not have 
% a background in your field.
%
% A Plain Language Summary is required in GRL, JGR: Planets, JGR: Biogeosciences,
% JGR: Oceans, G-Cubed, Reviews of Geophysics, and JAMES.
% see http://sharingscience.agu.org/creating-plain-language-summary/)
%
%% ------------------------------------------------------------------------ %%

%% \begin{abstract} starts the second page

\begin{abstract}

Being able to accurately predict the arrival of coronal mass ejections (CMEs) at Earth has been a long-standing problem in space weather research and operations. In this study, we use the ELlipse Evolution model based on Heliospheric Images (ELEvoHI) to predict the arrival time and speed of 10 CME events that were observed by HI on the STEREO-A spacecraft between 2010 and 2020. Additionally, we introduce a Python tool for downloading and preparing STEREO-HI data, as well as tracking CMEs. In contrast to most previous studies, we use not only science data, which have a relatively high spatial and temporal resolution, but also \rev{lower-quality} beacon data, which are --- in contrast to science data --- provided in real-time by the STEREO-A spacecraft. We do not use data from the STEREO-B spacecraft. We get a mean absolute error of 8.81 $\pm$ 3.18~h / 59 $\pm$ 31~km~s$^{-1}$ for arrival time/speed predictions using science data and 11.36 $\pm$ 8.69~h / 106 $\pm$ 61~km~s$^{-1}$ for beacon data. We find that using science data generally leads to more accurate predictions, but using beacon data with the ELEvoHI model is certainly a viable choice in the absence of higher resolution real-time data. We propose that these differences could be minimized if not eliminated altogether if higher quality real-time data were available, either by enhancing the quality of the already available data or coming from a new mission carrying a HI instrument on-board.

\end{abstract}

%TC:ignore
\section*{Plain Language Summary}

Coronal mass ejections (CMEs) are large ejections of plasma and the accompanying magnetic field caused by magnetic activity on the Sun. If CMEs reach Earth, they interact with the planetary magnetic field. In doing so, CMEs can cause disturbances to power grids and other electrical infrastructure on our planet, inhibit radio transmissions and damage satellites, which is why it is important to have an accurate way of predicting the arrival of the phenomena. Our model uses data provided by the HI cameras on the STEREO spacecraft. These data are available in a lower quality in real-time, i.e. within a latency of about 5 minutes within being received at the ground station, or in a higher quality with a delay of around 3 days. Using real-time data is important if we want to be able to predict the arrival of CMEs in a timely manner. In this study, we show that we can use the lower-quality real-time STEREO-HI data to make accurate predictions of the arrival time of CMEs.
%TC:endignore
%%% Suggested section heads:
\section{Introduction}\label{sec:Introduction}

\rev{Coronal mass ejections (CMEs)} are explosive outbursts of plasma and the accompanying magnetic field from the Sun. The expelled material moves at high speeds throughout the heliosphere and interacts with any obstacles, such as planets and their magnetospheres, along the way. If a CME reaches Earth, various phenomena, summarized under the term space weather, can be observed. These phenomena can range from harmless to destructive in their intensity. CMEs may induce strong geomagnetic storms that have a significant impact on satellites in orbit and electrical devices on the planet's surface as well as the ability to cause disturbances in radio transmissions. As our world becomes more and more reliant on technology and thus a continuous supply of electricity to power it, the importance of real-time ICME predictions is increasing~\citep{can13, oug17}.

A variety of modelling techniques have been developed for this purpose. In the following, we give a short overview focusing on models that rely on heliospheric imagers on board the STEREO spacecraft. Examples include Fixed-Phi Fitting \citep[FPF;][]{rou08, she99}, which assumes that the CME is a point moving radially away from the Sun at constant speed in a fixed direction and Harmonic Mean Fitting \citep[HMF;][]{lug09b, lug10}, which makes the same assumptions as FPF but treats the CME as an expanding sphere with the center tethered to the Sun. For Self-Similar Expansion Fitting \citep[SSEF;][]{dav12, moedav13}, the width of the CME may be freely chosen anywhere between 0$^\circ$, in which case it is identical with FPF, and 180$^\circ$, in which case it is equivalent to HMF.

While FPF, HMF and SSEF methods can provide an estimate of the CME’s propagation direction and speed based on single-spacecraft HI observations, multi-point views can be exploited to reduce the assumptions necessary. The “triangulation” class of methods assumes the same geometries used in FPF \citep{liu10a}, HMF \citep{lug10} and SSEF \citep{lug10, dav13}. As was recently shown by \citet{hin21}, the assumption of a fixed shape for CMEs can lead to differences in arrival time predictions depending on which spacecraft is used as imaging input within the model.

Slightly more sophisticated models may employ drag-based fitting. Such methods assume that the kinematics of a CME (which are dominated by the Lorentz force close to the Sun) are governed by the ambient solar wind flow as the transient moves away from the Sun. CMEs that are faster than the ambient solar wind are slowed down while slower ones speed up \citep{car04, vrs13}. A model building on this assumption is the Drag-based Ensemble Model (DBEM) first introduced by \citet{dum18}. It produces a range of outcomes for the arrival time and speed of the CME. The Ellipse Evolution Model based on Heliospheric Imager data \citep[ELEvoHI;][]{rol16, ame18, ame21} which is used in this paper is based on observations made using the Heliospheric Imagers \citep[HI;][]{eyl09} aboard the STEREO spacecraft. ELEvoHI incorporates drag-based model fitting and is able to give estimates for certain input parameters due to its use of HI observations. This method will be described in greater detail in \Cref{sec:Methods}.

Models for the prediction of CME parameters may also be empirical in nature, using parameters observed in situ to make predictions. One such model is the Effective Acceleration Model \citep[EAM;][]{pao17}, which relies on the SOHO/LASCO and ACE spacecraft. More computationally intense models rely upon the solution of the magneto-hydrodynamic (MHD) equations. \rev{These include, for example, ENLIL \citep{ods04} and EUHFORIA \citep{pom18}, which can be coupled with different coronal models, such as \citep[WSA;][]{arg03}, Magnetohydrodynamics Algorithm outside a Sphere \citep[MAS;][]{link99} and MULTI-VP \citep{pin17}.}

Models predicting the arrival time of CMEs are being improved continuously. To be of use for establishing a system that could give an advance warning of the arrival of an Earth directed CME, models must be able to predict such events in real-time. The \rev{prediction} models that are currently operational mostly make use of coronagraph data from the Large Angle and Spectrometric Coronagraph (LASCO) aboard the Solar and Heliospheric Observatory (SOHO), which sits at L1 and has a field of view (FOV) of about 30 R$_\odot$ in all directions. Data from the COR2 coronagraph aboard STEREO-A are also used frequently. There are already models predicting the arrival of CMEs in real time that have been operating for years, e.g. WSA-ENLIL+Cone \citep{ods04}. \citet{wol18} assessed the performance of the WSA-ENLIL+Cone model for 273 CMEs that occurred between March 2010 and December 2016 and found a mean absolute arrival time error of 10.4 ± 0.9 hours. \citet{moe17} analyzed 1337 CME events from a period of 8 years and predicted their arrival time using the SSE technique. This approach yielded an absolute arrival time error of 2.6 ± 16 h and found that prediction accuracy for STEREO-HI science data slightly increases with increasing longitudinal observer angle. There is, however, no current model that allows for real-time predictions using the STEREO-HI instruments. 

The STEREO-HI instruments benefit from a larger FOV compared to coronagraphs, which are centered on the Sun directly and mostly observe the solar corona. Additionally, \rev{STEREO-HI has the ability to view Earth-directed CMEs side-on when at or near the L4 or L5 orbital points}, instead of only head-on as a coronagraph would, resulting in a better view of their structure. These benefits could improve the accuracy of CME arrival predictions and decrease the number of false alarms. CME kinematics may change considerably due to interactions with the solar wind and/or other CMEs during their interplanetary propagation, making prolonged observation of the event advantageous \citep{col13}. Unfortunately, real-time data from the STEREO-HI instruments are only available in low-rate beacon data \rev{\citep{bie08}}. These \rev{lower-quality} data make real-time observation of a CME further away from the Sun more difficult.

\citet{tuc15} have previously used beacon data in combination with FPF as part of the Solar Storm Watch (SSW) citizen science project to analyze the effect that the use of real-time data have on predictions of arrival speed and transit time. For the 20 CMEs that arrived at Earth, they obtained an arrival speed error of 151~km~s$^{-1}$ and a transit time error of 22 h. By taking interplanetary acceleration of the CME into account, they were able to reduce the error to 77~km~s$^{-1}$ for the arrival speed and 19 h for the transit time. These errors were attributed to a mix of technical issues with the real-time data, such as the low spatial and temporal resolution, and physical issues with the propagation of CMEs. Furthermore, \citet{davi11} obtained an arrival time error of + 12.98 h for the 8 April 2010 event, which is also studied in this paper, using STEREO-B HI beacon data.

\rev{Using ELEvoHI, we expect to see results similar to those in previously mentioned studies in terms of arrival time and speed errors for predictions based on science data. Outcomes for predictions based on beacon data are expected to be worse than those based on science data. The general aim of space weather predictions is giving accurate arrival time and speed estimates for CMEs in real-time, so as to ensure a reliable way of alerting interested parties to an oncoming CME before it actually arrives. ELEvoHI is constantly being improved in terms of accuracy, most recently trough the incorporation of frontal deformation for CMEs by \citet{hin21b}. The implementation of predictions using beacon data for ELEvoHI would add to the tool's repertoire and be a step towards true real-time predictions of CME arrival time and speed.}

In \Cref{sec:Data}, we give an overview of the data we use and how they were prepared. In \Cref{sec:Methods}, we describe the basic components of the ELEvoHI model. In \Cref{sec:Results} we present our results and describe the differences between predictions made using STEREO-HI science images and the so-called beacon data (real-time data) in detail. The results are discussed in \Cref{sec:Discussion}. \Cref{sec:Summary} provides a summary and a conclusion of our work.

\section{Data}\label{sec:Data}

\subsection{The STEREO mission}

The twin-spacecraft STEREO-A(head) and STEREO-B(ehind) were launched in 2006 to improve our understanding of various space weather phenomena, including CMEs  and particularly those CMEs that are Earth directed \citep{kai08}. STEREO-A moves on an orbit inside the Earth's orbit around the Sun, while STEREO-B's orbit lies slightly further out. \rev{The difference in orbital distance is small, i.e. both spacecrafts' orbit close to 1 AU from the Sun, but large enough for STEREO-B to lag behind STEREO-A in its orbit around the Sun.} This leads to the two spacecraft attaining an increasing angular separation, which allows for observing space weather phenomena from two distinct vantage points. The focus of this paper lies on the use of STEREO-A data. \rev{Contact with STEREO-B was lost in 2014 after a test of the spacecraft's command loss timer before it entered into a period of solar conjunction.}

In this work, we make use of STEREO's Sun Earth Connection Coronal and Heliospheric Investigation \citep[SECCHI;][]{how08} suite of instruments, which consists of two white-light imagers with overlapping FOV, called heliospheric imagers \citep[HI1 and HI2;][]{eyl09} as well as two white-light coronagraphs, COR1 and COR2 \citep{tho03, how08}, and one EUV-camera \citep[EUVI;][]{wue04}. The FOV of the two HI instruments is measured in degrees of elongation. Elongation gives the angle between the observer, Sun-centre and another object. 0$^\circ$ of elongation corresponds to an object directly on the observer-Sun line. HI1 has a FOV extending from roughly 4$^{\circ}$ to 24$^{\circ}$ elongation when measured from the Sun center, giving it a FOV \rev{of} 20 x 20 $^{\circ}$. HI2 has a FOV extending from 18.8$^{\circ}$ to 88.8$^{\circ}$ elongation, amounting to a FOV of \rev{70 x 70} $^{\circ}$. These values are valid in the ecliptic plane during nominal operations. The time cadence differs between data types and instruments, with HI1 science data having a cadence of 40 minutes and HI1/HI2 beacon data as well as HI2 science data having a cadence of 120 minutes.

\begin{figure}
\includegraphics[width=\textwidth]{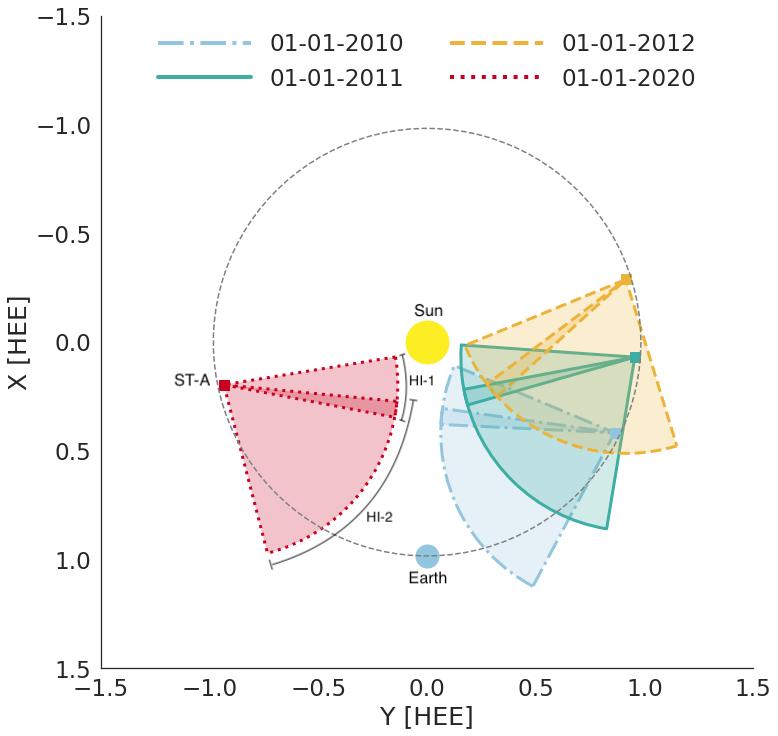}
\caption{Orbital position of the STEREO-A spacecraft on 1 January, 00:00:00 UT of each year in which an event under study took place. The region within the spacecrafts FOV where the HI1 and HI2 instruments overlap is indicated by a darker shade of the respective color. The position of STEREO-B is omitted in this figure since only data from STEREO-A are used. The blue dot represents Earth.}
\label{fig:STEREOOrbits}
\end{figure}

To test the viability of making real-time predictions using STEREO-HI beacon data with the ELEvoHI model, we analyze 10 CMEs observed by the STEREO-A spacecraft between 2010 and 2020. We chose the events because they have been extensively studied before (except for the 9 July 2020 event) and are particularly well visible in STEREO-HI science data \citep{moe14, rol16}. Furthermore, they encompass a variety of dates with data captured from different positions of STEREO-A's orbit. The orbital position of STEREO-A for each year in which one of the selected events occurs is depicted in \Cref{fig:STEREOOrbits}.

Data transmitted by STEREO's space weather beacon are of considerably lower quality than regular science data. Science data are transferred at regular intervals upon contact with the Deep Space Network (DSN) while beacon data are continually broadcast to a number of cooperating antenna stations around the world. Due to the limitations in telemetry allocated to the real-time beacon mode, HI2 beacon data undergo Rice lossless compression while HI1 data are compressed using ICER lossy compression. Furthermore, convolutional 1/6 encoding (changed from convolutional 1/2 encoding on 27 July 2007) is used to ensure reliable data transfer for beacon data and reduce errors in transmission. Both forms of data are uploaded in near-real time onto the internet after being downlinked and are subsequently processed into Flexible Image Transport System (FITS) files at the STEREO Science Center. The images are made available to the public as soon as this process is finished, which means that beacon data are available in near real-time \citep{eic08}.

\subsection{Data preparation}

The images used in this work go through extensive pre-processing to minimize any residual noise and make the CMEs as clearly visible as possible. Usually, data correction and calibration of STEREO/HI images is done in part by using the $\tt{secchi\_prep}$ routine, written in IDL$^{\rm TM}$ SolarSoft. The data reduction procedures contained therein were adapted for the Python programming language for this paper and are available online as open-source code (see \Cref{sec:DataCodes}). Using a freely available open-source programming language ensures that reproducibility of results is not contingent on access to software via an institutional subscription. A brief overview of the most important steps is given below.

Both science and beacon images are already processed aboard the STEREO spacecraft, but this will not be discussed further at this point. For a detailed description of the on-board processing, please refer to \citet{eyl09}. Science and beacon data are treated largely equally during the on-ground data reduction process. In the beginning, saturated columns are masked. Saturation often occurs when planets are in the imagers FOVs. Neither HI1 nor HI2 utilize camera shutters, so smearing effects that will occur due to this circumstance must be corrected. To be able to easily interpret the image data, a calibration factor can be applied that converts the units of the pixels from DN/sec to units of mean solar brightness. A flatfield is subtracted from the images and distortion correction is done to account for the HI cameras wide-angle optics.  As one of the last steps, the spacecraft's pointing information is updated by fitting the image to known stars.

In order to obtain the time-elongation profile necessary for making predictions with the ELEvoHI model, time-elongation-plots are created from the L1 processed images obtained from the Python program. Tracking can also be done directly in L1 HI images. This is sometimes the case for studies investigating individual CME events, as for example in \citet{rol14}. Producing a time-elongation map \citep[J-Map;][]{dav09b, she99} is accomplished by extracting a strip from the image with a fixed width of 5$^{\circ}$ in position angle centered on the ecliptic plane. This process is completed for each of the images in the time-series. The strips obtained this way are stacked next to each other, producing a J-Map. We use running difference images to further enhance the visibility of the CMEs under study. This produces the final J-Map in which the leading edge of the CME is clearly visible as a bright streak.

\section{Methods}\label{sec:Methods}

ELEvoHI was first introduced by \citet{rol16} as a combination of the ELlipse Evolution model \citep[ELEvo;][]{moe15} and drag-based model fitting (DBM fitting) that can be used to predict arrival time and speed of CMEs at various points within the heliosphere. The authors showed that the addition of the solar wind drag leads to improvements in CME prediction, particularly in terms of arrival speed, when compared to previously described methods such as FPF, HMF or SSEF \rev{\citep{rol16}}. ELEvoHI was first presented as a single-run model, but has since been updated to employ an ensemble approach \citep{ame18, ame21}.

\begin{figure}
\centering
\includegraphics[scale=0.55]{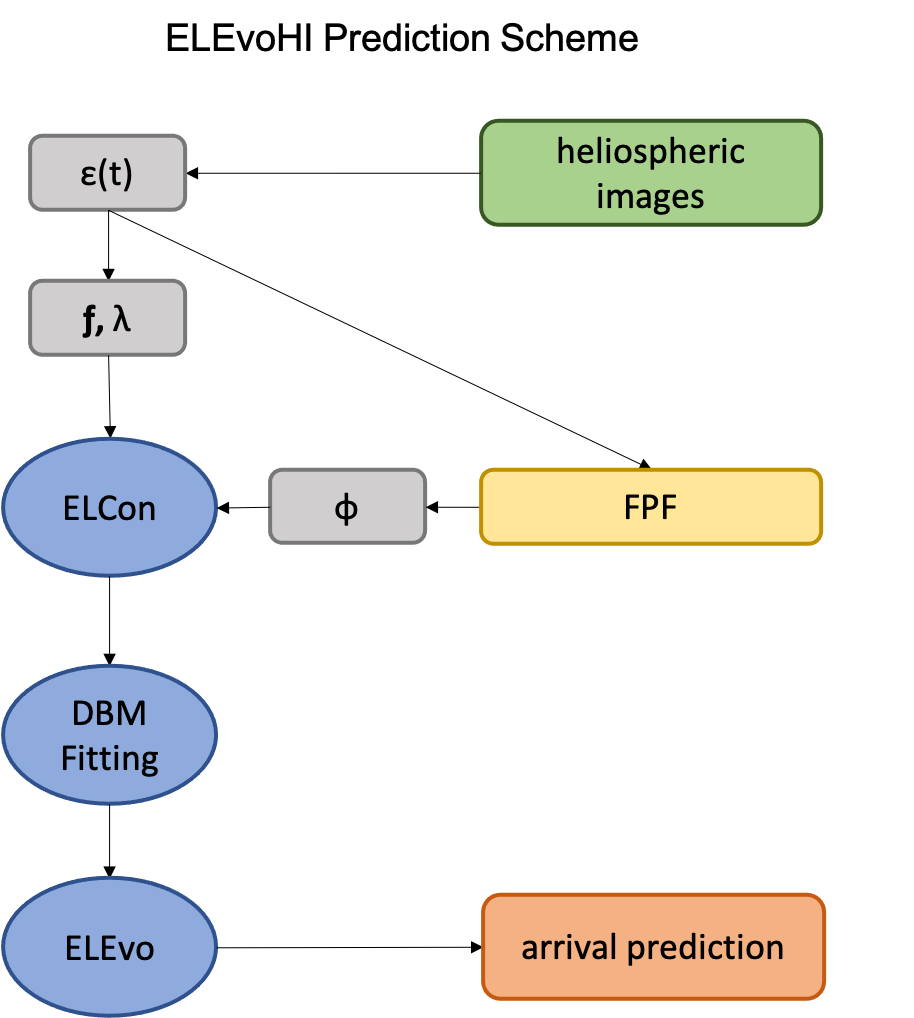}
\caption{Components and their connections within ELEvoHI. The gray boxes show the input parameters \rev{(the elongation $\epsilon{t}$, the inverse ellipse aspect ratio $f$, the half-width $\lambda$ and the direction of motion $\phi$)} and their source data (shown in a green box) as well as the different input sources (shown in a yellow box) which are used. The blue boxes signify ELEvoHI's three main models. The \rev{orange} box contains the models main output which is used in this work.}
\label{fig:ELEvoHI}
\end{figure}

ELEvoHI consists of three main methods, ELlipse Conversion (ELCon), DBM fitting and ELEvo. DBM fitting and ELCon provide parameters for ELEvo in order to generate predictions for CME arrival time and speed. \Cref{fig:ELEvoHI} presents an overview of how the modules within ELEvoHI relate to each other. A time-elongation track of the CME is required as input. Such a track is obtained by tracing the CME in J-Maps. \rev{To minimize the influence of human tracking deviations between tracks, we attempt to track each CME 5 times for each data type and event. These 5 tracks are then interpolated onto a common time grid via polynomial interpolation and averaged.} The elongation profile must be converted to units of radial distance. In the ELEvoHI model, this task is taken on by ELCon which works under the assumption of an elliptical geometry for the CME. Similarly to SSE, the CMEs angular half-width, $\lambda$, can be freely chosen. Additionally, the inverse ellipse aspect ratio, $f$, can be modified to change the curvature of the CME front. At $f$ = 1, the frontal shape is equal to that of a circle. As $f$ decreases, the front becomes flatter. The CME's direction of motion, $\phi$, must also be provided for ELCon to be able to accurately convert the elongation profile into a radial distance profile. \rev{ELCon is consistent with the assumption of an elliptical CME geometry and, in combination with ensemble modeling, allows for a variety of CME shapes to be considered for predictions.}

In this study, a range of values for $f$, $\phi$ and $\lambda$ are used. $f$ and $\lambda$ are estimated quantities in this study. Recently, \citet{hin21} introduced the Ecliptic cut Angles from Graduated Cylindrical Shell \citep{the06} for ELEvoHI \rev{tool (EAGEL)} which opens up the possibility of a better estimation of $\lambda$. We vary $f$ between 0.7 and 1.0 with a step size of 0.1. We vary $\phi$ over a range of $\pm$ 10$^{\circ}$ from the values acquired via the FPF method with a step size of 2$^{\circ}$. We vary $\lambda$ between 30 and 50, with a step size of 5. After the conversion via ELCon, the aforementioned parameters serve as input for DBM fitting, a fitting technique based on the equations of the DBM. Additionally, DBM fitting requires the selection of a range of R(t) to consider. These ranges are manually selected for each event. This method delivers the ambient solar wind speed $\omega$, the drag parameter $\gamma$, the initial radial distance $r_{\rm init}$ and speed $v_{\rm init}$ at time $t_{\rm init}$. $\omega$ is determined by inputting a range of background solar wind speeds (250 ~--~ 700~km~s$^{-1}$, in 25~km~s$^{-1}$ steps) and subsequently choosing the speed that yields the best DBM fit. These initial values are thus derived solely from the CME kinematics. The values are passed onto ELEvo, in combination with the angular half width and the inverse aspect ratio, which then predicts the arrival time and speed of the CME based on the assumption of an elliptical CME front with constant half-width and aspect ratio.

\rev{The selected CMEs are compared regarding the difference in predicted and in situ arrival time and speed. In situ arrival times for all events, except 9 July 2020, are taken from \citet{moe14}, which uses data from the Wind spacecraft (see \Cref{sec:DataCodes}) to determine arrival of a CME at L1 by using the shock arrival time as the CME arrival time. The in situ arrival time for the 9 July 2020 event was determined through similar means. The CME speed at L1 is also obtained from Wind data. It is taken to be the mean proton bulk speed of either of the aforementioned phenomena (that is the CME itself, or the shock ahead of the magnetic flux rope, MFR, or indeed the MFR itself).}

\section{Results}\label{sec:Results}

\begin{figure}
\includegraphics[width=\textwidth]{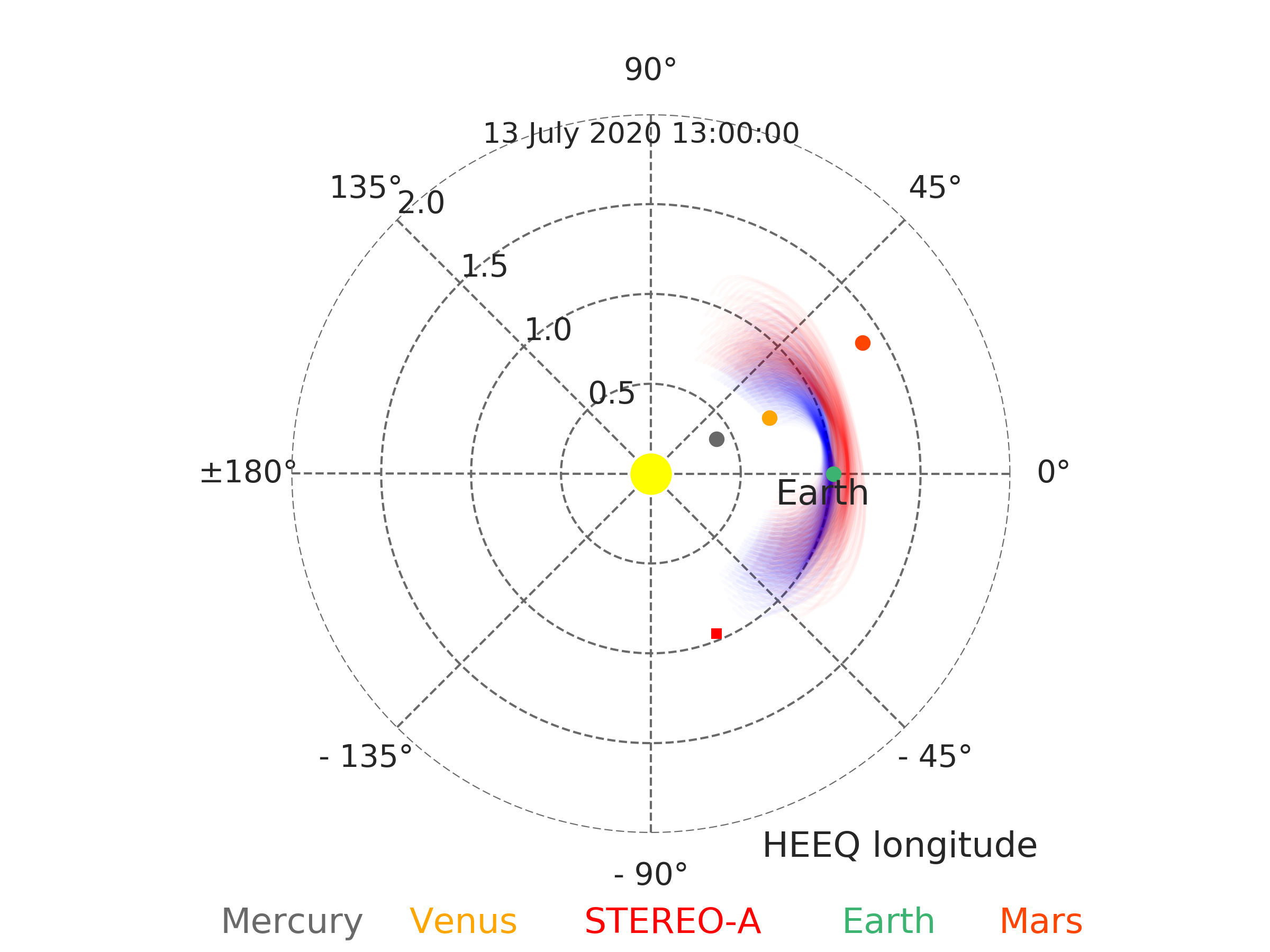}
\caption{Snapshot from a movie showing the results of the ELEvoHI ensemble run for the 9 July 2020 event. Both ensemble runs using science (blue) and beacon (red) tracks as starting points are shown. Earth is marked as a green dot, the position of STEREO-A is given by the red rectangle.}
\label{fig:MovieCut}
\end{figure}

We perform ELEvoHI ensemble modeling for 10 CME events observed between the years of 2010 and 2020 by the STEREO-A spacecraft and compare the predicted arrival times and speeds obtained using science and beacon data to in-situ arrivals. Each ensemble run consists of 275 individual ensemble members. All selected CMEs propagated in or close to the ecliptic plane, making them clearly visible from STEREO-A's perspective at the time of observation. The dates given for each event correspond to the date that the respective CME was first observed on in the STEREO-HI1 camera. \rev{\Cref{fig:MovieCut} shows a still image taken from a movie depicting the results of an ELEvoHI ensemble run for one of our selected events. Science data are marked in blue, while beacon data are shown in red. Each ensemble member is shown as an elliptical front moving outwards from the Sun. As this figure shows, there is variation within ensembles even for the same data type due to the use of differing CME parameters. It also shows the discrepancy between predictions based on science and beacon data, with ensemble members of this beacon data prediction generally arriving earlier than those  of science data predictions.}

\subsection{Arrival time and speed predictions}

\begin{figure}
\includegraphics[width=\textwidth]{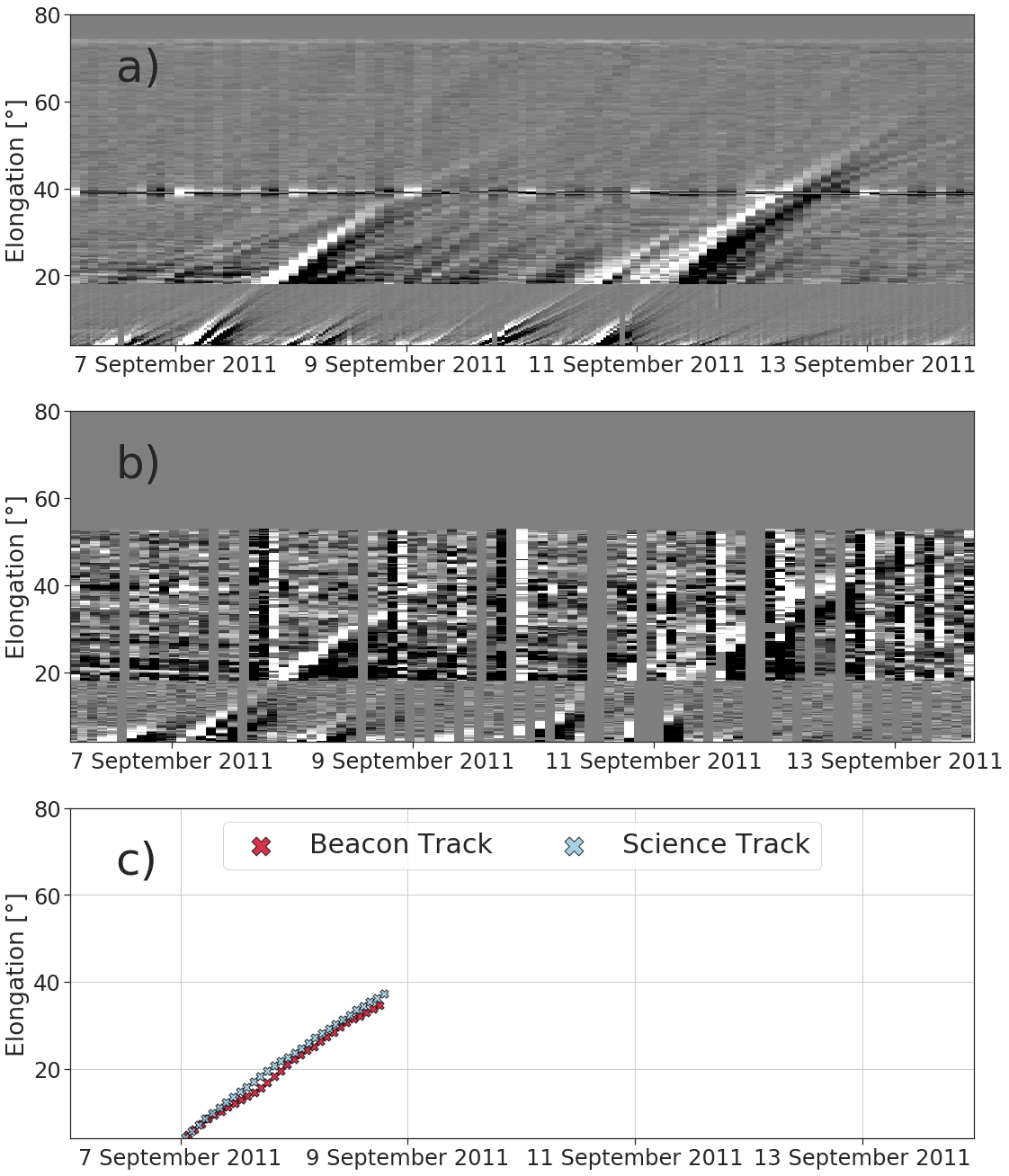}
\caption{A time-elongation map (J-Map) of the 7 September 2011 event \rev{showing both HI-1 and HI-2 data, separated by a distinct line at approximately 18.5$^{\circ}$}. Panel a) shows the J-Map made using science data. Panel b) shows the J-Map generated using beacon data. The gray vertical stripes in the beacon J-Map correspond to data gaps. Beacon data have a lower spatial and temporal resolution than science data. \rev{Panel c) shows the tracks for science and beacon data. The blue marks represent the track obtained via averaging the 5 individual tracks made of that event using the science J-Map. The red marks represent the corresponding beacon track.}}
\label{fig:JplotComp}
\end{figure}

As already described in the previous section, the time-elongation tracks are obtained by manually tracking the path of the CME front along the ecliptic in J-Maps resulting from the data reduction procedures. We do this 5 times for every event and data type \rev{(science and beacon)}; so there are a total of 10 tracks for each event. The 5 tracks of each data type are interpolated to lie along a regularly spaced time-axis and subsequently averaged to decrease the influence of human tracking errors on the result \rev{(more on this in \Cref{sec:trackerr})}. \rev{\Cref{fig:JplotComp} shows a comparison of J-Maps for science and beacon data. The J-Map produced using science data is clearly of superior quality, with much fewer data gaps and fainter structures also clearly visible.}

\begin{figure}
\includegraphics[width=\textwidth]{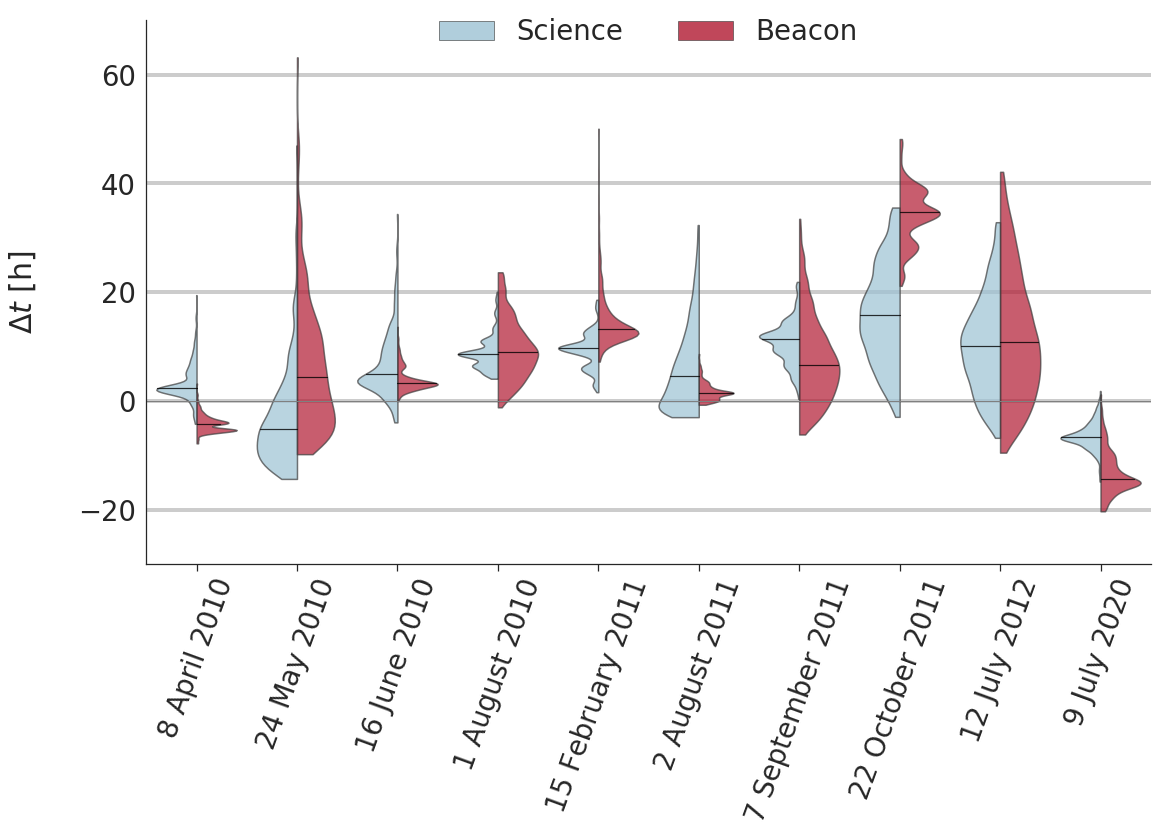}
\caption{Difference in predicted and in situ arrival time ($\Delta t$ = $t_{\rm ELEvoHI}$ - $t_{\rm obs}$) given in hours at Earth for all selected CME events. The left-hand side of the violin in blue represents the distribution of the $\Delta t$ for all ensemble members made using science data, the right-hand side in red represents the same distribution for beacon data. The horizontal black lines mark the median values of the respective distributions. Positive values indicate a late prediction, while negative values signify an early prediction.}
\label{fig:PredTime}
\end{figure}

\Cref{fig:PredTime} shows the distributions of the difference in predicted and in situ arrival time for all ten events as violin plots based on science (blue) and beacon (red) data. The error in hours for science data is given on the left-hand side of each violin in blue, while the error for beacon data for the same CME is given on the right-hand side in red. The dark horizontal line in each distribution marks the median value. Positive values indicate that the prediction made using ELEvoHI succeeds the actual in situ arrival, while negative values signify a premature prediction. The mean absolute error, MAE\rev{(t)}, of the arrival time predictions over all 10 CMEs made using science data is 8.8~h, the mean error, ME\rev{(t)}, is 6.2~h and the root mean square error, RMSE\rev{(t)}, is 8.9~h. The standard deviation, STD\rev{(t)}, of MAE\rev{(t)} for science data is 3.2~h. The MAE\rev{(t)} for beacon data is 11.4~h, the ME\rev{(t)} is 7.3~h and the RMSE\rev{(t)} is 13.9~h. The STD\rev{(t)} of MAE\rev{(t)} for beacon data is 8.7~h.

Looking at the ME\rev{(t)} for each event suggests that ELEvoHI tends to predict a late arrival for science as well as for beacon data, when compared to the actual in situ arrival time. There are 4 exceptions to this observation: The ME\rev{(t)} of the time-difference distribution obtained using beacon data of the 8 April 2010 CME, using science data of the 24 May 2010 CME and for both data types of the 9 July 2020 CME. All of these predictions are early, on average. The event which shows the largest discrepancy for both science and beacon data is 22 October 2011, with a deviation of 15.7~h for science and 34.7~h for beacon data, respectively. For science data, this is also the event possessing the largest standard deviation with a value of 8.9~h. For beacon data, the event with the largest standard deviation is 12 July 2012, with a value of 10.3~h. The ME\rev{(t)}s, MAE\rev{(t)}s and RMSE\rev{(t)}s for $\Delta t$, including each event's standard deviations can be found in \Cref{tab:delta_all}.

\begin{figure}
\includegraphics[width=\textwidth]{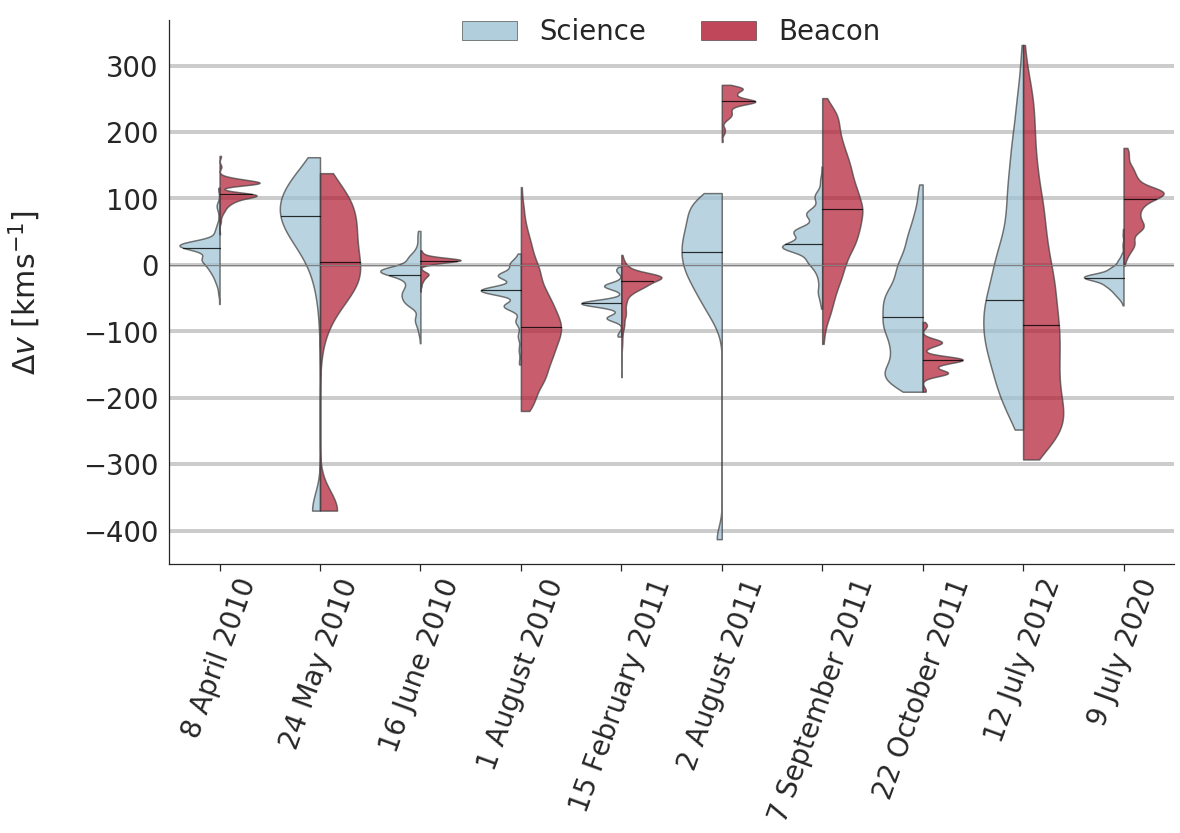}
\caption{\rev{Difference in predicted and in situ arrival speed ($\Delta v$ = $v_{\rm ELEvoHI}$ - $v_{\rm obs}$) given in km~s$^{-1}$ at Earth for all selected CME events. The left-hand side of the violin in blue represents the distribution of the $\Delta v$ for all ensemble members made using science data, the right-hand side in red represents the same distribution for beacon data. The horizontal black lines mark the median values of the respective distributions. Positive values indicate a prediction slower than the in situ arrival speed, while negative values signify a prediction faster than the in situ measurements.}}
\label{fig:PredSpeed}
\end{figure}

\Cref{fig:PredSpeed} shows the distributions of the difference in predicted and in situ arrival speed at L1 for all ten events as violin plots based on science (blue) and beacon (red) data. The mean error in km~s$^{-1}$ for science data \rev{is given on} the left-hand side of the violins in blue, beacon data is given on the right-hand side in red. The dark horizontal line in each distribution indicates the median value. Positive values signify that the speed predicted by ELEvoHI is higher than the actual in situ speed, while negative values indicate that the predicted speed was lower than that of the actual CME. The mean absolute error, MAE(v), of the arrival speed predictions made using science data is 59~km~s$^{-1}$, the mean error, ME(v), is -14~km~s$^{-1}$ and the root mean square error, RMSE(v), is 39~km~s$^{-1}$. The standard deviation of MAE(v), STD(v), is 31~km~s$^{-1}$. The MAE(v) for beacon data is 106~km~s$^{-1}$, the ME(v) is 17~km~s$^{-1}$ and the RMSE(v) is 111~km~s$^{-1}$. The STD(v) of MAE(v) is 61~km~s$^{-1}$.

A particular bias in the predictions is less clear for speed predictions than it is for arrival time predictions. The ME(v)s indicate that ELEvoHI tends to predict an arrival speed that is slower than observed for science data and faster than observed for beacon data. It must be noted, however, that the numbers for the beacon data arrival speed are significantly influenced by the 2 August 2011 event which has an arrival speed error distribution that skews heavily towards a speed prediction that is too fast. The 2 August 2011 event has an MAE(v) of 246~km~s$^{-1}$, making it the event with the largest MAE(v) in terms of arrival speed of any event using beacon data. The largest MAE(v) for arrival speed for science data belongs to the 12 July 2012 event with 104~km~s$^{-1}$. The event which shows the largest standard deviation is 24 May 2010, for both science and beacon data, with a deviation of 91~km~s$^{-1}$ and 115~km~s$^{-1}$, respectively. The MAE(v)s including their standard deviations for all events can be found in \Cref{tab:delta_all}.

%%%%%%%%%%%%%%%%%%%%%%%%%%%%%%%%%%%%%%%%%%%%%%%%%%%%%%%%%%%%%%%%%%%%%%%%%%%%%%%%%%%%%%%%%%
\begin{table}
  \centering
  \renewcommand{\arraystretch}{1.2}
\begin{adjustbox}{center}
\begin{tabularx}{1.24\textwidth}{lrrrrrrrr}
    \toprule
    \multirow{2}{*}{\textbf{Events}} & \multicolumn{2}{c}{\textbf{ME\rev{(t)} [h]}} & \multicolumn{2}{c}{\textbf{MAE\rev{(t)} [h]}} & \multicolumn{2}{c}{\textbf{RMSE\rev{(t)} [h]}} & \multicolumn{2}{c}{\textbf{STD\rev{(t)} [h]}}\\
    % \hline
    % \textbf{Inactive Modes} & \textbf{Description}\\
    \cline{2-9}
    & \textbf{Science} & \textbf{Beacon} & \textbf{Science} & \textbf{Beacon} & \textbf{Science} & \textbf{Beacon} & \textbf{Science} & \textbf{Beacon}\\
    %\hhline{~--}
    \midrule
    8 April 2010 &       3.3 &      -4.2 &        3.7 &        4.3 &         4.8 &         4.5 &        3.1 &        1.4 \\
    24 May 2010 &      -3.0 &       6.6 &        7.7 &       10.0 &         9.6 &        14.0 &        5.8 &        9.8 \\
    16 June 2010 &       6.7 &       3.9 &        7.0 &        3.9 &         9.1 &         4.5 &        5.8 &        2.2 \\
    1 August 2010 &       8.9 &       9.5 &        8.9 &        9.5 &         9.4 &        11.0 &        3.2 &        5.5 \\
    15 February 2011 &       9.3 &      14.9 &        9.3 &       14.9 &         9.9 &        15.8 &        3.2 &        5.3 \\
    2 August 2011 &       6.1 &       1.7 &        7.0 &        1.8 &         9.9 &         2.4 &        6.9 &        1.6 \\
    7 September 2011 &      10.8 &       7.6 &       10.8 &        8.5 &        11.5 &        11.0 &        3.9 &        7.0 \\
    22 October 2011 &      15.9 &      34.3 &       16.0 &       34.3 &        18.3 &        34.6 &        8.9 &        5.1 \\
    12 July 2012 &      10.5 &      11.8 &       11.2 &       13.1 &        13.7 &        16.6 &        8.0 &       10.3 \\
    9 July 2020 &      -6.5 &     -13.3 &        6.5 &       13.3 &         6.9 &        13.8 &        2.5 &        3.9 \\
    \bottomrule
    \\
    \multirow{2}{*}{\textbf{Events}} & \multicolumn{2}{c}{\textbf{ME(v) [km~s$^{-1}$]}} & \multicolumn{2}{c}{\textbf{MAE(v) [km~s$^{-1}$]}} & \multicolumn{2}{c}{\textbf{RMSE(v) [km~s$^{-1}$]}} & \multicolumn{2}{c}{\textbf{STD(v) [km~s$^{-1}$]}}\\
    % \hline
    % \textbf{Inactive Modes} & \textbf{Description}\\
    \cline{2-9}
    & \textbf{Science} & \textbf{Beacon} & \textbf{Science} & \textbf{Beacon} & \textbf{Science} & \textbf{Beacon} & \textbf{Science} & \textbf{Beacon}\\
    %\hhline{~--}
    \midrule
    8 April 2010 &               22 &              110 &                27 &               110 &                 34 &                112 &                21 &                17 \\
    24 May 2010 &               36 &              -32 &               105 &                98 &                139 &                152 &                91 &               115 \\
    16 June 2010 &              -24 &                2 &                30 &                10 &                 39 &                 12 &                25 &                 6 \\
    1 August 2010 &              -41 &              -89 &                42 &                98 &                 52 &                113 &                31 &                57 \\
    15 February 2011 &              -56 &              -33 &                56 &                33 &                 60 &                 42 &                22 &                26 \\
    2 August 2011 &                5 &              245 &                60 &               245 &                 96 &                246 &                75 &                17 \\
    7 September 2011 &               38 &               84 &                44 &                96 &                 53 &                115 &                29 &                64 \\
    22 October 2011 &              -73 &             -142 &                89 &               142 &                105 &                144 &                55 &                21 \\
    12 July 2012 &              -32 &              -65 &               111 &               141 &                132 &                164 &                71 &                84 \\
     9 July 2020 &              -19 &               92 &                22 &                92 &                 26 &                100 &                13 &                37 \\
    \bottomrule
\end{tabularx}
\end{adjustbox}
  \caption{The ME, MAE and RMSE of the difference between in situ and predicted arrival time $\Delta t$ = $t_{\rm ELEvoHI}$ - $t_{\rm obs}$ and speed $\Delta v$ = $v_{\rm ELEvoHI}$ - $v_{\rm obs}$ at Earth for each of the events under study. The standard deviation for each respective quantity and event and is also given. \rev{The MAE(t) of all predictions based on science data is 8.8~h, the ME(t) is 6.2~h and the RMSE(t) is 8.9~h. The STD(t) of the MAE(t) for science data is 3.2~h. The MAE(t) for predictions based on beacon data is 11.4~h, the ME(t) is 7.3~h and the RMSE(t) is 13.9~h. The STD(t) of the MAE(t) for beacon data is 8.7~h. The mean absolute error, MAE(v) of predictions made using science data is 59~km~s$^{-1}$, the ME(v) is -14~km~s$^{-1}$ and the the RMSE(v) is 39~km~s$^{-1}$. STD(v) of MAE(v) is 31~km~s$^{-1}$. The MAE(v) for predictions made using beacon data is 106~km~s$^{-1}$, the ME(v) is 17~km~s$^{-1}$ and the RMSE(v) is 111~km~s$^{-1}$. The STD(v) of MAE(v) is 61~km~s$^{-1}$ for beacon data.}}
\label{tab:delta_all}
\end{table}

%%%%%%%%%%%%%%%%%%%%%%%%%%%%%%%%%%%%%%%%%%%%%%%%%%%%%%%%%%%%%%%%%%%%%%%%%%%%%%%%%%%%%%%%%%

\subsection{Influence of human tracking error}
\label{sec:trackerr}

It is important to keep in mind that the time-elongation profiles describing the CME's trajectory outward are obtained by tracing the perceived front of the CME by hand. The process is thus subject to human error. Different people may come up with disparate tracks in the end, even if guidelines are provided. \citet{bar15} analyzed and compared three different methods for tracking CMEs (CMEs were tracked by experts, an algorithm or participants in the SSW citizen science program) and found that the method used introduced more variability into the assumed CME kinematics than the differences between several single-spacecraft fitting techniques did. Even if only one person is tasked with tracking the front, as was the case in this study, there is no guarantee that tracks will always be identical, although the difference between tracks generated by the same individual may be smaller than that of tracks coming from several people.

\begin{figure}
\includegraphics[width=\textwidth]{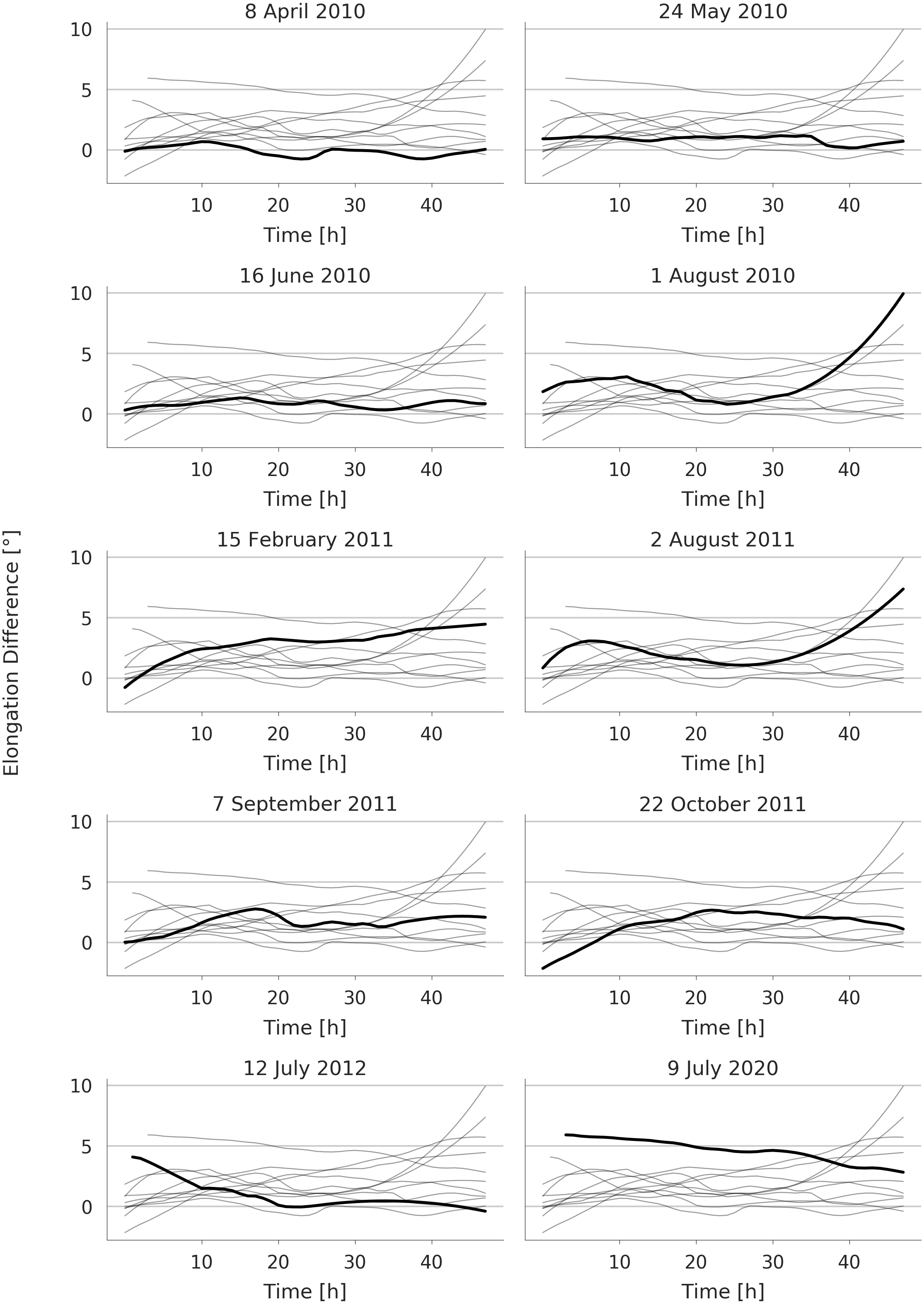}
\caption{The average difference in elongation between science and beacon tracks for all events under study using averaged tracks. The difference at each point in time is taken and plotted against the time steps on the time-grid. Marked as thicker line are the elongation differences for the particular event, the fainter gray lines in the background are the elongation differences of all other events meant to serve as a comparison.}
\label{fig:ElonDiff}
\end{figure}

\Cref{fig:ElonDiff} shows the difference between the averaged science and beacon time-elongation profiles for each event. The difference between the science and beacon tracks for each event at each point in time was calculated and is plotted against the time-axis. The thick line marks the difference in elongation for each event, while the fainter gray lines in the background represent the difference in elongation for all other dates which serve as a comparison. The event with the largest average difference between science and beacon data track is 9 July 2020. In contrast, the two tracks are almost identical for the 8 April 2010 event. \rev{On average, science data tracks have a larger elongation at the same point in time than their beacon data counterparts. There are 4 events (8 April 2010, 15 February 2011, 22 October 2011 and 12 July 2012) where tracks obtained from beacon data show a larger elongation \rev{for at least one point in time} than tracks obtained using science data.}

To investigate the influence that the deviations between individual tracks have on the overall outcome of the time and speed predictions, we input all 5 tracks for each data type and event separately into ELEvoHI and see how the results differ from each other. To be able to ascertain the differences between tracks for the same event, we look at \rev{the difference between predicted and in situ measurements of arrival time, $\Delta t$, and arrival speed, $\Delta v$.} For the beacon data tracks of event 2 August 2011, no predictions from individual tracks based on science or beacon mode were possible, suggesting that this event benefited greatly from the interpolation and averaging of the 5 separate tracks. Furthermore, this event already displayed quite a large error in its beacon data arrival speed prediction over the ensemble members. \rev{Sometimes, tracks may simply not have enough data points or be too irregularly spaced, preventing ElEvoHI from making a prediction based on that track. Badly chosen starting points for the DBM fit or a poorly tracked time-elongation profile may also cause failed ELEvoHI predictions.}

\begin{figure}
\includegraphics[width=\textwidth]{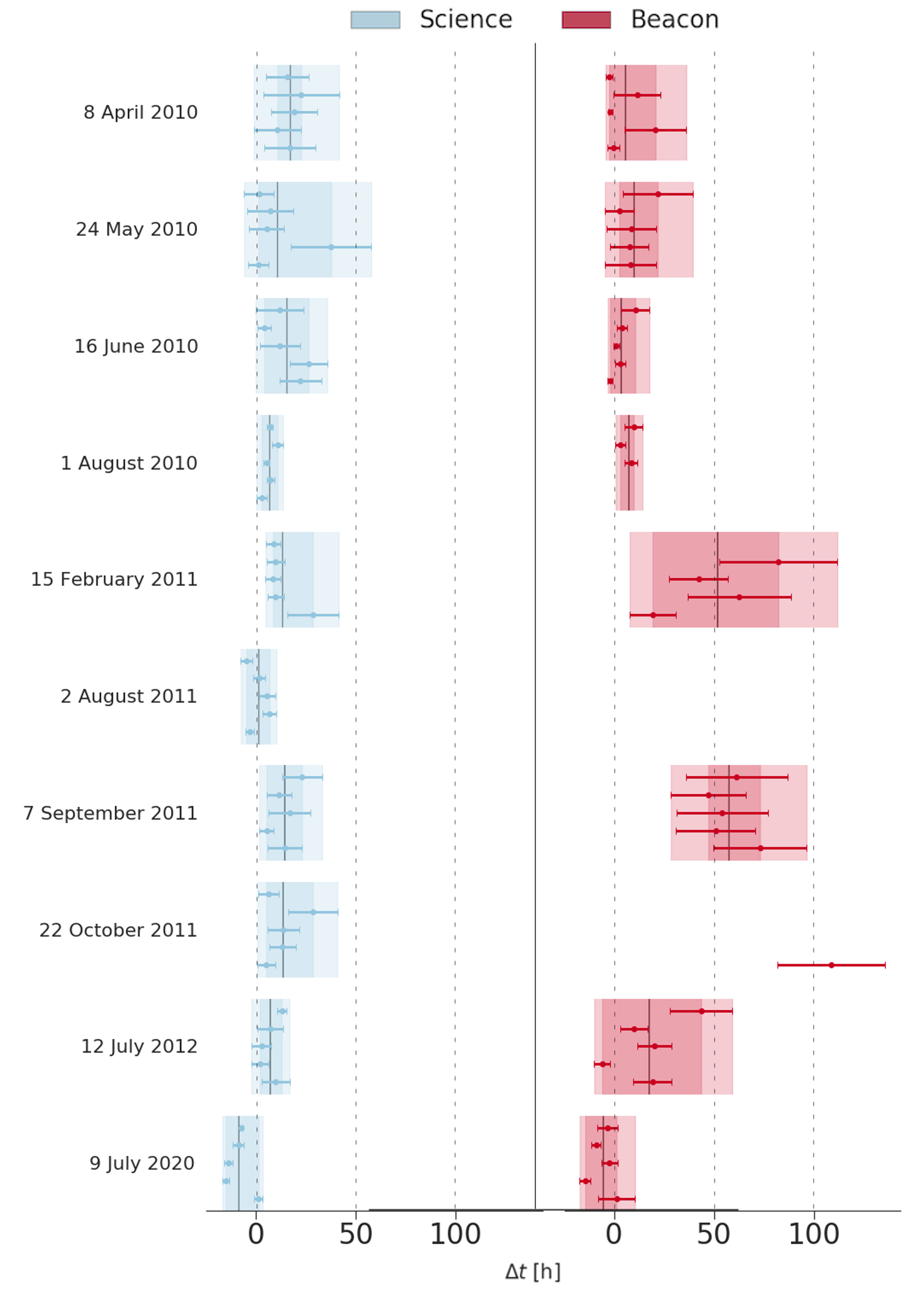}
\caption{Plots showing $\Delta t$ in hours for each event under study for science and beacon data. Instead of using the average track obtained by interpolating/averaging all 5 tracks for each event and data type, the 5 individual tracks are input into ELEvoHI separately and the outcomes are summarized here. The blue (left) part of the plot shows the results for all science tracks of each respective event, while the red (right) part shows the results for all beacon tracks. Note that a prediction using any individual beacon data tracks for the 2 August 2011 event was not possible. The shaded regions show the maximum range of $\Delta t$, Range($\Delta t$) = ME(t)$_{\rm max}$ - ME(t)$_{\rm min}$, for each respective date and data type. Each dot represents the mean of the prediction for one track; the bars represent a standard deviation of 1 $\sigma$. The gray vertical line across each shaded region represents the mean of all 5 tracks.}
\label{fig:dt_single}
\end{figure}

\Cref{fig:dt_single} shows the distribution of $\Delta t$ for all ten CMEs in terms of their mean and standard deviation. $\Delta t$ in hours for science data is always given on the left-hand side in blue \rev{while it is given on the right-hand side marked in red for beacon data}. The average MAE(t) of the predictions made using science data is 11.8~h. The average MAE(t) for predictions made using beacon data is 23.5~h. Science data have, on average, a considerably lower MAE(t) than beacon data. The largest MAE(t) for any event tracked in science data is 37.7~h obtained for event 24 May 2010; for beacon data it is event 22 October 2011 with a MAE(t) of 108.9~h.

\begin{figure}
\includegraphics[width=\textwidth]{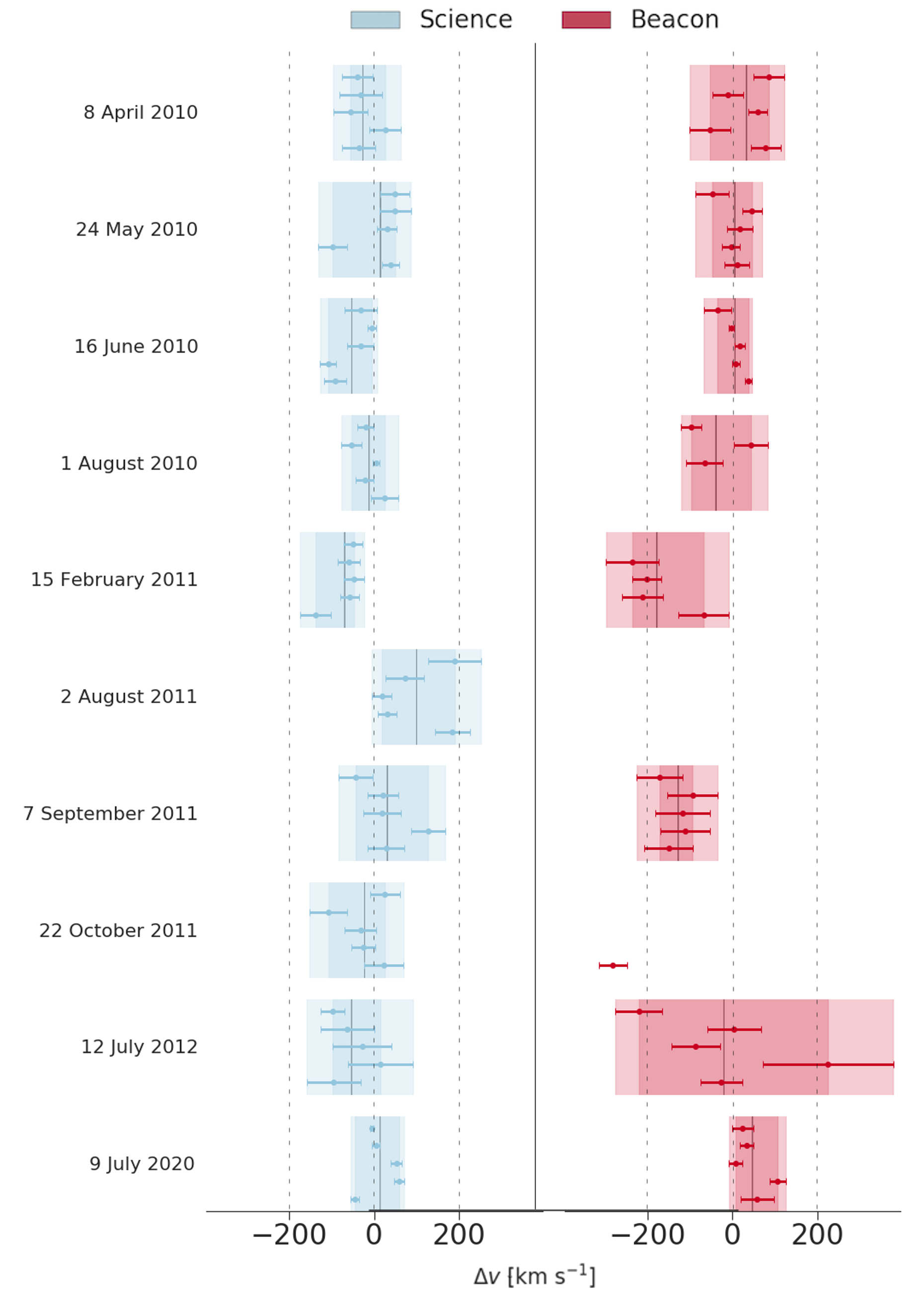}
\caption{\rev{Plots showing $\Delta v$ in km~s$^{-1}$ for each event under study for science and beacon data. 5 individual, non-averaged tracks are input into ELEvoHI separately and the outcomes are compared here. The blue (left) part of the plot shows the results for all science tracks of each respective event, while the red (right) part shows the results for all beacon tracks. A prediction using any individual beacon data tracks for the 2 August 2011 event was not possible. The shaded regions show the maximum range of $\Delta v$, Range($\Delta v$) = ME(v)$_{\rm max}$ - ME(v)$_{\rm min}$, for each respective date and data type. Each dot represents the mean of the prediction for one track; the bars indicates a standard deviation of 1 $\sigma$. The gray vertical line across each shaded region represents the mean of all 5 tracks.}}
\label{fig:dv_single}
\end{figure}

\Cref{fig:dv_single} shows the same quantities as \Cref{fig:dt_single} for $\Delta v$. The average MAE(v) of the predictions made using science data is 66~km~s$^{-1}$. The average MAE(v) for predictions made using beacon data is 93~km~s$^{-1}$. The largest MAE(v) for any individual event tracked in science data is 189~km~s$^{-1}$ obtained for the 2 August 2011 event; for beacon data it is event 22 October 2011 with a MAE(v) of 281~km~s$^{-1}$. All results pertaining to the $\Delta t$ and $\Delta v$ can be found in \Cref{tab:dt_dv_single} for science data and beacon data.

%%%%%%%%%%%%%%%%%%%%%%%%%%%%%%%%%%%%%%%%%%%%%%%%%%%%%%%%%%%%%%%%%%%%%%%%%%%%%%%%%%%%%%%%%%
\aboverulesep = 0.1ex
\belowrulesep = 0.1ex

\begin{table}[p]
\centering
\begin{adjustbox}{center}
\begin{tabularx}{1.32\textwidth}{llrrrrrrrr}
%TC:ignore
\toprule
    \multirow{2}{*}{\textbf{Events}} & \multirow{2}{*}{\textbf{Track}} & \multicolumn{2}{c}{\textbf{MAE\rev{(t)} [h]}} & \multicolumn{2}{c}{\textbf{STD\rev{(t)} [h]}} & \multicolumn{2}{c}{\textbf{MAE(v) [km~s$^{-1}$]}} & \multicolumn{2}{c}{\textbf{STD(v) [km~s$^{-1}$]}}\\
    \cline{3-10}
    & & \textbf{Science} & \textbf{Beacon} & \textbf{Science} & \textbf{Beacon} & \textbf{Science} & \textbf{Beacon} & \textbf{Science} & \textbf{Beacon}\\
\midrule
 8 April 2010 &         1 &           16.9 &            3.8 &           12.7 &            3.0 &                     44&                     82&                     40&                     35\\
     8 April 2010 &         2 &           11.3 &           20.8 &           11.8 &           15.4 &                     64&                     58&                     37&                     47\\
     8 April 2010 &         3 &           19.1 &            2.2 &           11.5 &            1.1 &                     57&                     60&                     40&                     21\\
     8 April 2010 &         4 &           22.7 &           11.8 &           19.0 &           11.8 &                     68&                     44&                     50&                     36\\
     8 April 2010 &         5 &           15.6 &            3.3 &           10.8 &            1.8 &                     43&                     86&                     36&                     36\\
     \midrule
      24 May 2010 &         1 &            5.1 &           10.3 &            5.2 &           12.8 &                     42&                     48&                     20&                     30\\
      24 May 2010 &         2 &           37.7 &            8.5 &           20.1 &            9.6 &                     98&                     27&                     34&                     21\\
      24 May 2010 &         3 &            8.2 &           10.9 &            8.8 &           12.5 &                     43&                     51&                     24&                     29\\
      24 May 2010 &         4 &           10.8 &            6.3 &           11.6 &            7.2 &                     64&                     53&                     36&                     24\\
      24 May 2010 &         5 &            7.9 &           21.9 &            7.4 &           17.6 &                     57&                     54&                     34&                     40\\
      \midrule
     16 June 2010 &         1 &           22.2 &            2.1 &           10.4 &            1.2 &                     91&                     38&                     26&                      8\\
     16 June 2010 &         2 &           26.3 &            3.4 &            9.4 &            2.7 &                    108&                     13&                     19&                      9\\
     16 June 2010 &         3 &           12.0 &            2.3 &           10.1 &            1.5 &                     38&                     19&                     32&                     12\\
     16 June 2010 &         4 &            4.1 &            3.9 &            3.4 &            2.5 &                     14&                     12&                     10&                      6\\
     16 June 2010 &         5 &           11.8 &           10.6 &           12.0 &            7.1 &                     41&                     39&                     39&                     32\\
     \midrule
    1 August 2010 &         1 &            3.1 &            - &            2.7 &            - &                     53&                      - &                     32&                      - \\
    1 August 2010 &         2 &            7.3 &            - &            2.0 &            - &                     25&                      - &                     20&                      - \\
    1 August 2010 &         3 &            5.2 &            8.6 &            1.4 &            3.2 &                     16&                     71&                      8&                     43\\
    1 August 2010 &         4 &           10.8 &            3.1 &            2.8 &            2.4 &                     53&                     65&                     24&                     39\\
    1 August 2010 &         5 &            7.0 &            9.8 &            1.6 &            4.4 &                     22&                     97&                     20&                     24\\
    \midrule
 15 February 2011 &         1 &           28.5 &           19.3 &           12.9 &           11.5 &                    138&                     70&                     37&                     59\\
 15 February 2011 &         2 &            9.9 &           62.7 &            4.1 &           25.8 &                     58&                    211&                     22&                     48\\
 15 February 2011 &         3 &            8.5 &           42.3 &            3.8 &           14.7 &                     47&                    201&                     24&                     35\\
 15 February 2011 &         4 &            9.8 &           82.3 &            4.6 &           29.6 &                     59&                    235&                     26&                     62\\
 15 February 2011 &         5 &            8.7 &            - &            3.6 &            - &                     50&                      - &                     22&                      - \\
 \midrule
    2 August 2011 &         1 &            4.2 &            - &            2.1 &            - &                    184&                      - &                     41&                      - \\
    2 August 2011 &         2 &            6.7 &            - &            3.6 &            - &                     43&                      - &                     22&                      - \\
    2 August 2011 &         3 &            5.6 &            - &            4.1 &            - &                     43&                      - &                     24&                      - \\
    2 August 2011 &         4 &            3.7 &            - &            3.0 &            - &                     74&                      - &                     45&                      - \\
    2 August 2011 &         5 &            5.7 &            - &            2.9 &            - &                    189&                      - &                     62&                      - \\
    \midrule
 7 September 2011 &         1 &           14.4 &           73.1 &            8.6 &           23.4 &                     68&                    150&                     44&                     57\\
 7 September 2011 &         2 &            5.3 &           50.9 &            3.6 &           19.9 &                    126&                    111&                     40&                     59\\
 7 September 2011 &         3 &           16.8 &           54.1 &           10.4 &           22.8 &                     68&                    117&                     43&                     64\\
 7 September 2011 &         4 &           11.5 &           47.2 &            6.3 &           18.8 &                     51&                     95&                     35&                     59\\
 7 September 2011 &         5 &           23.0 &           61.4 &           10.1 &           25.5 &                     49&                    171&                     40&                     54\\
 \midrule
  22 October 2011 &         1 &            6.5 &          108.9 &            4.5 &           26.9 &                     64&                    281&                     46&                     32\\
  22 October 2011 &         2 &           13.2 &            - &            6.6 &            - &                     41&                      - &                     28&                      - \\
  22 October 2011 &         3 &           13.6 &            - &            7.9 &            - &                     58&                      - &                     37&                      - \\
  22 October 2011 &         4 &           28.5 &            - &           12.3 &            - &                    108&                      - &                     45&                      - \\
  22 October 2011 &         5 &            7.1 &            - &            5.1 &            - &                     52&                      - &                     36&                      - \\
  \midrule
     12 July 2012 &         1 &           10.6 &           19.2 &            6.9 &            9.7 &                    116&                     64&                     63&                     48\\
     12 July 2012 &         2 &            6.1 &            7.0 &            4.3 &            4.0 &                    103&                    233&                     77&                    153\\
     12 July 2012 &         3 &            6.7 &           20.1 &            5.0 &            8.5 &                    106&                     94&                     69&                     57\\
     12 July 2012 &         4 &            8.7 &           10.0 &            6.4 &            6.9 &                    107&                    102&                     63&                     63\\
     12 July 2012 &         5 &           12.9 &           43.6 &            2.5 &           15.6 &                     98&                    220&                     29&                     55\\
     \midrule
      9 July 2020 &         1 &            1.5 &           10.3 &            2.1 &            9.3 &                     46&                     70&                     10&                     39\\
      9 July 2020 &         2 &           15.2 &           14.5 &            1.7 &            2.8 &                     58&                    106&                     11&                     19\\
      9 July 2020 &         3 &           14.0 &            6.3 &            2.1 &            4.1 &                     52&                     26&                     13&                     16\\
      9 July 2020 &         4 &            8.9 &            9.1 &            2.7 &            2.5 &                     15&                     34&                      9&                     16\\
      9 July 2020 &         5 &            7.4 &            7.8 &            1.0 &            5.3 &                      8&                     38&                      4&                     25\\
\bottomrule
\end{tabularx}
\end{adjustbox}
%TC:endignore
\caption{The MAE and STD of the difference between in situ and predicted arrival time $\Delta t$ = $t_{\rm ELEvoHI}$ - $t_{\rm obs}$ and speed $\Delta v$ = $v_{\rm ELEvoHI}$ - $v_{\rm obs}$ at Earth for science and beacon data for each of the events under study. The quantities are given for each of the 5 tracks separately.}
\label{tab:dt_dv_single}
\end{table} 
%%%%%%%%%%%%%%%%%%%%%%%%%%%%%%%%%%%%%%%%%%%%%%%%%%%%%%%%%%%%%%%%%%%%%%%%%%%%%%%%%%%%%%%%%%

To further highlight the difference between predictions made using different tracks, \Cref{tab:range_single} lists the largest differences between arrival time and speed predictions, defined as Range($\Delta t$) = ME(t)$_{\rm max}$ - ME(t)$_{\rm min}$ and Range($\Delta v$) = ME(v)$_{\rm max}$ - ME(v)$_{\rm min}$, for the same event and data type. Over all CMEs, the average Range($\Delta t$) is 17.9~h for science data and 24~h for beacon data while the average Range($\Delta v$) is 119~km~s$^{-1}$ for science and 137~km~s$^{-1}$ for beacon data. The maximum Range($\Delta t$) is 36.5~h for the 24 May 2010 event for science data and 63.0~h for the 15 February 2011 event for beacon data. For Range($\Delta v$), the maximum is 172~km~s$^{-1}$ for the 2 August 2011 event for science data and 444~km~s$^{-1}$ for the 12 July 2012 event for beacon data.

%%%%%%%%%%%%%%%%%%%%%%%%%%%%%%%%%%%%%%%%%%%%%%%%%%%%%%%%%%%%%%%%%%%%%%%%%%%%%%%%%%%%%%%%%%
\begin{table}[ht]
\centering
\begin{adjustbox}{center}
\begin{tabularx}{0.74\textwidth}{lrrrr}
%TC:ignore
\toprule
\multirow{2}{*}{\textbf{Event}} & \multicolumn{2}{c}{\textbf{ Range($\Delta t$) [h]}} & \multicolumn{2}{c}{\textbf{Range($\Delta v$) [km~s$^{-1}$]}}\\
    \cline{2-5}
    & \textbf{Science} & \textbf{Beacon} & \textbf{Science} & \textbf{Beacon}\\
\midrule
     8 April 2010 &                                     11.9 &                                     23.3 &                                     82&                                    139\\
      24 May 2010 &                                     36.5 &                                     19.2 &                                    147&                                     94\\
     16 June 2010 &                                     22.2 &                                     12.6 &                                    103&                                     74\\
    1 August 2010 &                                      8.0&                                      6.7 &                                     78&                                    140\\
 15 February 2011 &                                     20.0 &                                     63.0 &                                     91&                                    168\\
    2 August 2011 &                                     11.5 &                                      - &                                    172&                                      - \\
 7 September 2011 &                                     17.8 &                                     25.9 &                                    170&                                     77\\
  22 October 2011 &                                     23.4 &                                      0.0 &                                    133&                                      0\\
     12 July 2012 &                                     11.0 &                                     49.6 &                                    112&                                    444\\
      9 July 2020 &                                     16.5 &                                     15.7 &                                    104&                                     99\\
\bottomrule 
\end{tabularx}
\end{adjustbox}
%TC:endignore
\caption{The maximum of the range of values of all tracks for $\Delta t$, Range($\Delta t$) = ME(t)$_{\rm max}$ - ME(t)$_{\rm min}$, and $\Delta v$, Range($\Delta v$) = ME(v)$_{\rm min}$ - ME(v)$_{\rm max}$ for each event and data type.}
\label{tab:range_single}
\end{table} 

%%%%%%%%%%%%%%%%%%%%%%%%%%%%%%%%%%%%%%%%%%%%%%%%%%%%%%%%%%%%%%%%%%%%%%%%%%%%%%%%%%%%%%%%%%
\section{Discussion}\label{sec:Discussion}

Our results show that the ELEvoHI arrival prediction based on HI science data is, on average, closer to the in situ arrival time and speed than the arrival prediction based on HI beacon data. However, it is not always the case that the science data predictions are closer to the in-situ arrival time and speed. There are several exceptions in terms of arrival time (events 24 May 2010, 16 June 2010, 2 August 2011 and 7 September 2011) as well as speed (events 24 May 2010, 16 June 2010 and 15 February 2011). The events studied in this work are unusually well visible in science and beacon data, which may lead to better prediction results for both data types. The number of events to choose from is also limited, as we included only Earth-directed CMEs in this study. Furthermore, all pre-2020 events are well-known CMEs investigated numerous times before. Familiarity with these events, as was the case for the CMEs in this paper, can certainly lead to anomalously good results that might not be possible if the event was tracked for the first time, especially without having seen the corresponding science data before. In order to further improve ELEvoHI's ability to make real-time predictions, it may be necessary to improve upon the current beacon data quality to be able to use data from a greater number of CMEs. However, it is shown that predictions for select well visible events using beacon data are viable and comparable to those made using science data.

The overall errors obtained for the 10 events under study are similar to the mean absolute errors found by other studies, although the results are sometimes not directly comparable since we perform hindcasting only while others predict CMEs which have, at the time that the prediction is made, not yet arrived at Earth. \rev{A mean absolute error of 8.6 $\pm$ 12.2~h was obtained by \citet{ame21} when running ELEvoHI in the same model setup for 15 events.} The current state of the art in terms of CME predictions was summarized by \citet{ril18}. In their analysis, they considered 32 models which were used to submit CME predictions to the Community Coordinated Modeling Center (CCMC) from 2013 to 2018. This online scoreboard is a way for researchers to compare their models in terms of arrival time\rev{.} They focused in particular on 28 events that were all predicted by 6 of the 32 models and found that these models were generally able to predict CME arrival times to within about $\pm$ 10~h, but that standard deviations often exceeded 20~h. The best performing model, the WSA-ENLIL+Cone model already described previously, achieved a mean absolute error of 13~h and a standard deviation of 15~h. 

Since averaging and interpolation of tracks aim to reduce the effects of human errors on the prediction, we also examined all events without applying these methods to qualitatively assess the effects that variations between tracks have on the prediction results. The standard deviations in elongation between all 5 tracks for each data type and event were rather low, with a mean absolute deviation of 0.22$^{\circ}$ for science data and 0.26$^{\circ}$ for beacon data. This suggests that if CMEs are tracked without a major time gap in between different attempts, tracks generated by the same person are very similar. To better understand the influence of these deviations within an ensemble and on the prediction results, we examined $\Delta t$ and $\Delta v$, as well as Range($\Delta t$) and Range($\Delta v$), for all 100 (10 events, 5 tracks per data type, 2 data types) non-averaged tracks. \Cref{fig:dt_single} and \Cref{fig:dv_single} show clearly that results of ensemble runs can differ greatly between different tracks for the same data type and event, even though the difference between tracks is rather small. Neither the error in arrival time and speed prediction nor the standard deviation within individual ensemble runs showed any clear correlation with the observer longitude of the STEREO-A spacecraft.

\section{Summary and Conclusion}\label{sec:Summary}
 
We used the ELEvoHI model to predict the arrival time and speed of 10 Earth directed CMEs between the years 2010 and 2020. ELEvoHI is commonly used with STEREO-HI science data, which have a higher spatial and temporal resolution but are not available in real-time, to make predictions. The model has the capability to deliver near-real time predictions for arrival time and speed if beacon data, which are downlinked in near-real time, are available for the desired time frame. In this work, we attempt to assess the feasibility of using beacon data with ELEvoHI, since it was unclear how viable predictions made with beacon data would be due to the low-quality nature of the data. Ensuring the possibility of real-time predictions using ELEvoHI is of great interest since STEREO-A is currently in an ideal position, and will be until around mid-2022, for observing the Sun-Earth line and thus possible Earth directed CMEs.

Each of the 10 CMEs selected for further study in this paper is a well-known event which is easily visible in science, and by extension also in beacon data. Each event was tracked in \rev{an} ecliptic J-Map by hand a total of 10 times, 5 times using J-Maps generated from science and 5 times using J-Maps generated from beacon data. The tracks for each event and data type were subsequently interpolated onto a regular time-grid and averaged to minimize the influence of any slight variations in the individual tracks. The ELEvoHI model with varying inverse ellipse aspect ratio $f$, direction of motion $\phi$ and angular half-width $\lambda$ is used to determine a distribution of arrival times and speeds. Each event is predicted once using science and once using beacon data. The median of both time and speed predictions, as well as the standard deviations for each distribution can be found in \Cref{tab:delta_all}. For predictions using science data, we obtained a MAE(t) of 8.8~h for the arrival time. The MAE(t) for beacon data was 11.4~h. In terms of speed, the MAE(v) for the science predictions amounted to 59~km~s$^{-1}$. For beacon data, the MAE(v) amounted to 106~km~s$^{-1}$.

We also input the 5 tracks for each event into ELEvoHI separately, without prior interpolation/averaging. For the MAE(t) of all runs made using science data, we found a mean value of 11.8~h. The mean MAE(t) for predictions made using beacon data is 23.5~h. The mean MAE(v) is 66~km~s$^{-1}$ for predictions made using science data and 93~km~s$^{-1}$ for predictions made using beacon data. A large variation in results between tracks of the same data type and date was observed for some dates, despite the fact that the tracks did not deviate from each other significantly. The tracks themselves have a mean absolute elongation deviation of 0.22$^{\circ}$ for science data and 0.26$^{\circ}$ for beacon data. The largest difference in $\Delta t$ for ensemble members of the same event is 36.5~h for the 24 May 2010 event for science data and 63.0~h for the 15 February 2011 event for beacon data. The largest Range($\Delta v$) for science data is 172~km~s$^{-1}$ for the 2 August 2011 event and 444~km~s$^{-1}$ for the 12 July 2012 event for beacon data.

\rev{We conclude that the availability of higher quality real-time data could possibly greatly improve the real-time predictions of CMEs using ELEvoHI or other HI-based methods. This could be achieved by launching a spacecraft carrying HI devices which is equipped for higher telemetry rates. The Lagrange point L5 would provide a perfect vantage point for the observation of Earth directed CMEs as it allows for continuous observation along the Sun-Earth line, unlike the STEREO spacecrafts' heliocentric drifting orbit. A spacecraft mission to said Lagrange point would also supply us with another point of view which could be combined with that of other spacecraft to improve our understanding of the evolution of the geometry of CMEs and possibly lead to more accurate solar wind forecasts \citep{sim09}.} Therefore, ESA's Lagrange mission to L5 could be a great step forward to improve the state of space weather modeling \citep{kra17}. Furthermore, with its planned launch into a low Earth orbit in 2023, the Polarimeter to UNify the Corona and Heliosphere (PUNCH) mission will also carry wide-angle HI cameras on board, enabling us to consider a near-Earth vantage point to increase the accuracy of our CME arrival predictions. \rev{As \citet{ame18} have shown, ELEvoHI is capable of making predictions for an HI observer that is located at the in-situ impact location.} \rev{However, in the absence of high quality real-time data from better equipped space missions, beacon data from the STEREO-A spacecraft have proven to be usable for real-time predictions. Improving upon the quality of the data already available might also be a viable path towards better real-time CME predictions.}

\section{Sources of Data, Codes and Supporting Information}\label{sec:DataCodes}

Data reduction program: 

https://github.com/helioforecast/STEREO-HI-Data-Processing/releases/tag/v1.0.0
(DOI: 10.5281/zenodo.5092136)
  
Animation of \Cref{fig:MovieCut}:

https://doi.org/10.6084/m9.figshare.14994345

In-situ data:

STEREO: https://stereo-ssc.nascom.nasa.gov/pub/

Wind: https://cdaweb.gsfc.nasa.gov/index.html/

\acknowledgments
T.A., J.H., M.B., M.R., C.M., A.J.W., and U.V.A. thank the Austrian Science Fund (FWF): P31265-N27, P31659-N27, P31521-N27.

\bibliography{main}

\end{document}